\documentclass[10pt,twocolumn]{article}
\usepackage{multirow}
\usepackage{fullpage}
\usepackage{graphicx}
\usepackage{times}
\usepackage{amssymb}
\usepackage{amsmath}

\newtheorem{theorem}{Theorem}[section]
\newtheorem{lemma}[theorem]{Lemma}

\newtheorem{definition}[theorem]{Definition}
\newtheorem{proposition}[theorem]{Proposition}

\newenvironment{example}{\refstepcounter{theorem}\smallskip\par\noindent{\bf Example \thetheorem.}}
{~\qed\normalfont\par\smallskip}

\newcommand{\qed}{$\Box$}
\newenvironment{proof}{\smallskip\par {\sc Proof.}}{~\qed\normalfont\par\smallskip}

\newcommand{\M}{{\cal M}}

\newcommand{\IdM}{\operatorname{Id}}
\newcommand{\IdF}{\overline{\IdM}}
\newcommand{\So}{{\mathbf S}}
\newcommand{\Ta}{{\mathbf T}}

\newcommand{\R}{{\mathbf R}}
\newcommand{\x}{\bar x}
\newcommand{\y}{\bar y}

\newcommand{\inst}{\operatorname{Inst}}

\newcommand{\certain}{\underline{\operatorname{certain}}}

\newcommand{\Sol}{\operatorname{Sol}}

\newcommand{\comp}{\circ}
\renewcommand{\subsetneq}{\varsubsetneq}

\newcommand{\dom}{\operatorname{dom}}

\newcommand{\pc}{\mathbf{C}}
\newcommand{\C}{\mathbf{C}}
\newcommand{\D}{\mathbf{D}}
\newcommand{\N}{\mathbf{N}}

\newcommand{\cq}{\text{\sc CQ}}
\newcommand{\ucq}{\text{\sc UCQ}}

\newcommand{\LL}{{\cal L}}
\newcommand{\cqa}[1]{\cq^{#1}}
\newcommand{\ucqa}[1]{\ucq^{#1}}

\newcommand{\tu}[2]{#1\text{\sc -to-}#2}

\newcommand{\cqcq}{\tu{\cq}{\cq}}

\newcommand{\ignore}[1]{}
\newcommand{\aproof}[2]{\noindent {\sc Proof of #1.}~#2\hfill{\qed}}

\newcommand{\sig}{{\rm sig}}
\newcommand{\at}{{\rm at}}
\newcommand{\bigtimes}{\mathop{\mathchoice%
{\raise-0.22em\hbox{\huge $\times$}}%
{\raise-0.05em\hbox{\Large $\times$}}{\hbox{\large $\times$}}{\times}}}

\newcommand{\cwa}{\operatorname{cws}}
\newcommand{\sk}{\operatorname{CQ-SkSTD}}
\newcommand{\sks}{\operatorname{CQ-SkSTDs}}
\newcommand{\compp}{\text{\sc Composition}}

\begin{document}

\title{{\bf Composition and Inversion of Schema
Mappings}\thanks{{\bf Database Principles Column.} Column editor:
Leonid Libkin, School of Informatics, University of Edinburgh,
Edinburgh, EH8 9AB, UK. E-mail: libkin@inf.ed.ac.uk.}}

\author{
\begin{tabular}{ccccccc}
Marcelo Arenas && Jorge P\'erez && Juan Reutter &&
Cristian Riveros\\  
PUC Chile && PUC Chile && U. of Edinburgh  && Oxford University\\
{\small marenas@ing.puc.cl} &&
{\small jperez@ing.puc.cl} &&
{\small juan.reutter@ed.ac.uk} &&
{\small cristian.riveros@comlab.ox.ac.uk}
\end{tabular}
}
\date{}        
\maketitle

\thispagestyle{empty}

\section{Introduction}
\label{sec-intro}


A schema mapping is a specification that describes how data from a
source schema is to be mapped to a target schema.  Schema mappings
have proved to be essential for data-interoperability tasks such as
data exchange and data integration. The research on this area has
mainly focused on performing these tasks. However, as Bernstein
pointed out~\cite{B03b}, many information-system problems involve not
only the design and integration of complex application artifacts, but
also their subsequent manipulation.  Driven by this consideration,
Bernstein proposed in ~\cite{B03b} a general framework for managing
schema mappings.  In this framework, mappings are usually
specified in a logical language, and high-level algebraic operators
are used to manipulate them
\cite{B03b,composing,Melnik04,inverse,BM07}.

Two of the most fundamental operators in this framework are the
\emph{composition} and \emph{inversion} of schema mappings.
Intuitively, the composition can be described as follows.  Given a
mapping $\M_1$ from a schema $\mathbf A$ to a schema $\mathbf B$, and
a mapping $\M_2$ from $\mathbf B$ to a schema $\mathbf E$, \emph{the
composition} of $\M_1$ and $\M_2$ is a new mapping that describes the
relationship between schemas $\mathbf A$ and $\mathbf E$.  This new
mapping must be
\emph{semantically consistent} with the relationships previously established
by $\M_1$ and $\M_2$.  On the other hand, \emph{an inverse} of $\M_1$
is a new mapping that describes the \emph{reverse} relationship from
$\mathbf B$ to $\mathbf A$, and is semantically consistent with
$\M_1$.  

In practical scenarios, the composition and inversion of schema
mappings can have several applications.  In a data exchange
context~\cite{de}, if a mapping $\M$ is used to exchange data from a
source to a target schema, an inverse of $\M$ can be used to exchange
the data back to the source, thus \emph{reversing} the application of
$\M$.  As a second application, consider a peer-data management system
(PDMS)~\cite{peer,piazza}.  In a PDMS, a peer can act as a data
source, a mediator, or both, and the system relates peers by
establishing \emph{directional} mappings between the peers schemas.
Given a query formulated on a particular peer, the PDMS must proceed
to retrieve the answers by reformulating the query using
its complex net of semantic mappings.  Performing this reformulation
at query time may be quite expensive.  The composition operator can be
used to essentially combine sequences of mappings into a single
mapping that can be precomputed and optimized for query answering
purposes. Another application is schema evolution,
where the inverse together with the composition play a crucial
role~\cite{BM07}.  Consider a mapping $\M$ between schemas $\mathbf A$
and $\mathbf B$, and assume that schema $\mathbf A$ evolves into a
schema $\mathbf A'$.  This evolution can be expressed as a mapping
$\M'$ between $\mathbf A$ and $\mathbf A'$.  Thus, the relationship
between the new schema $\mathbf A'$ and schema $\mathbf B$ can be
obtained by inverting mapping $\M'$ and then composing the result with
mapping $\M$.

In the recent years, a lot of attention has been paid to the development
of solid foundations for the composition~\cite{MH03,composing,NBM07}
and inversion~\cite{inverse,quasi,APR08,APRR09} of schema mappings.
In this paper, we review the proposals for the semantics of these
crucial operators.  For each of these proposals, we concentrate on the
three following problems: the definition of the semantics of the
operator, the language needed to express the operator, and the
algorithmic issues associated to the problem of computing the
operator.  It should be pointed out that we primarily consider the
formalization of schema mappings introduced in the work on data
exchange~\cite{de}.  In particular, when studying the problem of
computing the composition and inverse of a schema mapping, we will be
mostly interested in computing these operators for mappings specified
by
\emph{source-to-target tuple-generating dependencies} \cite{de}.
Although there has been an important amount of work about different
\emph{flavors} of composition and inversion motivated by practical
applications~\cite{BGMN08,MAB06,TBM09}, we focus on the most
theoretically-oriented
results~\cite{MH03,composing,inverse,quasi,APR08,APRR09}.

{\bf Organization of the paper.} We begin in Section 2 with the terminology that will be used in the paper.  
We then continue in Section 3 reviewing the main results for the composition operator 
proposed in~\cite{composing}.
Section~4 contains a detailed study of the inverse operators proposed in~\cite{inverse,quasi,APR08}. 
In Section 5, we review a relaxed approach to define the semantics 
for the inverse and composition operators that parameterizes these notions
by a query-language~\cite{MH03,APRR09}.
Finally, some future work is pointed out in Section 6, and the proofs of
the new results presented in this survey are given in Appendix
\ref{sec-app}. 

\section{Basic notation}
\label{sec-prel}
\noindent

In this paper, we assume that data is represented in the relational
model. A {\em relational schema} $\R$, or just {\em schema}, is a
finite set $\{R_1,\dots,R_n\}$ of relation symbols, with each $R_i$
having a fixed arity $n_i$. An instance $I$ of $\R$ assigns to each
relation symbol $R_i$ of $\R$ a finite $n_i$-ary relation $R^I_i$. The
{\em domain} of an instance $I$, denoted by $\dom(I)$, is the set of
all elements that occur in any of the relations $R_i^I$.  In addition,
$\inst(\R)$ is defined to be the set of all instances of $\R$.

As usual in the data exchange literature, we consider database
instances with two types of values: {\em constants} and {\em
nulls}. 
More precisely, let $\C$ and $\N$ be infinite
and disjoint sets of constants and nulls, respectively.
If we refer to a schema $\So$ as a {\em
source} schema, then $\inst(\So)$ is defined to be the set of all
instances of $\So$ that are constructed by using only elements from
$\C$, and if we refer to a schema $\Ta$ as a {\em target} schema, then
instances of $\Ta$ are constructed by using elements 
from both $\C$ and $\N$.

\medskip 
\noindent
{\bf Schema mappings and solutions.}  Schema mappings are used to
define a semantic relationship between two schemas. In this paper, we
use a general representation of mappings; given two schemas $\R_1$ and
$\R_2$, a mapping $\M$ from $\R_1$ to $\R_2$ is a set of pairs
$(I,J)$, where $I$ is an instance of $\R_1$, and $J$ is an instance of
$\R_2$. Further, we say that $J$ is a {\em solution for $I$ under
$\M$} if $(I,J) \in \M$. 
The set of solutions for $I$ under $\M$ is denoted by $\Sol_\M(I)$. 
The domain of $\M$, denoted by $\dom(\M)$, is defined as the set of
instances $I$ such that $\Sol_{\M}(I)\neq \emptyset$.

\medskip 
\noindent
{\bf Dependencies.} 
As usual, we use a class of {dependencies} 
to specify schema mappings \cite{de}.
Let $\LL_1$, $\LL_2$ be query languages and
$\R_1$, $\R_2$ be schemas with no relation symbols in common. A
sentence $\Phi$ over $\R_1 \cup \R_2$ is an
$\tu{\LL_1}{\LL_2}$ {\em dependency from $\R_1$ to $\R_2$} if $\Phi$
is of the form $\forall \x \, (\varphi(\x) \to \psi(\x))$, where (1)
$\x$ is the tuple of free variables in both $\varphi(\x)$ and
$\psi(\x)$; (2) $\varphi(\x)$ is an $\LL_1$-formula over $\R_1$; 
and (3) $\psi(\x)$ is an $\LL_2$-formula over $\R_2$.  
Furthermore, we usually omit the outermost universal
quantifiers from $\tu{\LL_1}{\LL_2}$ dependencies and, thus, we write
$\varphi(\x) \to \psi(\x)$ instead of $\forall \x \ (\varphi(\x) \to
\psi(\x))$. Finally, the semantics of an $\tu{\LL_1}{\LL_2}$ dependency
is defined as usual (e.g., see \cite{de,APR08}).

If $\So$ is a source schema and $\Ta$ is a target schema, an $\tu{\LL_1}{\LL_2}$ dependency
from $\So$ to $\Ta$ is called an $\tu{\LL_1}{\LL_2}$
\emph{source-to-target dependency} ($\tu{\LL_1}{\LL_2}$
st-dependency), and an $\tu{\LL_1}{\LL_2}$ dependency from $\Ta$ to
$\So$ is called an $\tu{\LL_1}{\LL_2}$ \emph{target-to-source
dependency} ($\tu{\LL_1}{\LL_2}$ ts-dependency). 
Notice that the fundamental class of source-to-target tuple-generating dependencies
(st-tgds)~\cite{de} corresponds to the class of $\cqcq$ st-dependencies.


When considering a mapping specified by a set of dependencies, we use
the usual semantics given by logical satisfaction.  That is, if $\M$
is a mapping from $\R_1$ to $\R_2$ specified by a set $\Sigma$ of
$\tu{\LL_1}{\LL_2}$ dependencies, we have that $(I,J)\in \M$ if and
only if $I\in \inst(\R_1)$, $J\in \inst(\R_2)$, and $(I,J)$ satisfies
$\Sigma$.

\smallskip
\noindent
{\bf Query Answering.} In this paper, we use $\cq$ to denote the class
of conjunctive queries and $\ucq$ to denote the class of unions of
conjunctive queries. 
Given a query $Q$ and a database instance $I$, we denote
by $Q(I)$ the evaluation of $Q$ over $I$. Moreover, we use
predicate $\pc(\cdot)$ to differentiate between constants and nulls,
that is, $\pc(a)$ holds if and only if $a$ is a constant value.
We use $=$, $\neq$, and $\pc$ as superscripts to denote
a class of queries enriched with equalities, inequalities, and
predicate $\pc(\cdot)$, respectively.
Thus, for example, $\ucq^{=,\pc}$ is the class of unions of conjunctive queries
with equalities and predicate $\pc(\cdot)$.

As usual, the semantics of queries in the presence of schema mappings
is defined in terms of the notion of {\em certain answer}. Assume that
$\M$ is a mapping from a schema $\R_1$ to a schema $\R_2$. Then given
an instance $I$ of $\R_1$ and a query $Q$ over $\R_2$, the {\em
certain answers of $Q$ for $I$ under $\M$}, denoted by
$\certain_\M(Q,I)$, is the set of tuples that belong to the evaluation
of $Q$ over every possible solution for $I$ under $\M$, that is,
$\bigcap \{ Q(J) \mid J$ is a solution for $I$ under $\M\}$.

\smallskip
\noindent
{\bf Proviso.} In this survey, only finite sets of dependencies are
considered.

\section{Composition of Schema Mappings}\label{sec-neg-cl}
\label{sec-comp}
\noindent


\noindent
The composition operator has been identified as one of the fundamental
operators for the development of a framework for managing schema
mappings \cite{B03b,Melnik04,MBHR05}. The goal of this operator is to
generate a mapping $\M_{13}$ that has the same effect as applying
successively two given mappings $\M_{12}$ and $\M_{23}$, provided that
the target schema of $\M_{12}$ is the same as the source schema of
$\M_{23}$. In
\cite{composing}, Fagin et al.~study the composition for the widely
used class of st-tgds. In particular, they provide solutions to the
three fundamental problems for mapping operators considered in this
paper, that is, they provide a formal semantics for the composition
operator, they identify a mapping language that is appropriate for
expressing this operator, and they study the complexity of composing
schema mappings. In this section, we present these solutions.

In \cite{composing,Melnik04}, the authors propose a semantics for the
composition operator that is based on the semantics of this operator
for binary relations:
\begin{definition}[\cite{composing,Melnik04}]
Let $\M_{12}$ be a mapping from a schema $\R_1$ to a schema
$\R_2$, and $\M_{23}$ a mapping from $\R_2$ to a schema $\R_3$. Then
the composition  of $\M_{12}$ and $\M_{23}$ is defined as 
$\M_{12} \circ \M_{23} = \{(I_1,I_3) \mid \exists I_2: (I_1,I_2) \in
\M_{12} \text{ and } (I_2,I_3) \in \M_{23} \}$. 
\end{definition}
Then Fagin et al.~consider in \cite{composing} the natural question of whether
the composition of two mappings specified by st-tgds can also be
specified by a set of these dependencies. 
Unfortunately, they prove in \cite{composing} that this is not the
case, as shown in the following example.

\newcommand{\takes}{\text{\tt Takes}}
\newcommand{\student}{\text{\tt Student}}
\newcommand{\enrollment}{\text{\tt Enrollment}}

\begin{example}{\bf \ (from \cite{composing})} \label{exa-tgs-compo}
Consider a schema $\R_1$ consisting of one binary relation \takes,
that associates a student name with a course she/he is taking, a
schema $\R_2$ consisting of a relation \takes$_\text{\tt 1}$, that is
intended to be a copy of \takes, and of an additional relation symbol
\student, that associates a student with a student id; and
a schema $\R_3$ consisting of a binary relation symbol \enrollment,
that associates a student id with the courses this student is taking.  Consider now
mappings $\M_{12}$ and $\M_{23}$ specified by the following sets
of st-tgds:
\begin{eqnarray*}
\Sigma_{12}  & = & \{ \takes(n,c) \to \takes_\text{\tt 1}(n,c),\\
&& \phantom{\{} \takes(n,c) \to \exists s \, \student(n,s) \},\\
\Sigma_{23}  & = & \{ \student(n,s) \wedge \takes_\text{\tt 1}(n,c) \ \to\\
&& \hspace{30mm} \enrollment(s,c) \}.
\end{eqnarray*}
\noindent
Mapping $\M_{12}$ requires that a copy of every tuple in \takes\ must
exist in \takes$_\text{\tt 1}$ and, moreover, that each student name
$n$ must be associated with some student id $s$ in the relation
\student. Mapping $\M_{23}$ requires that if a student with name $n$
 and id $s$ takes a course $c$, then $(s,c)$ is a tuple in the
relation \enrollment. Intuitively, in the composition mapping one
would like to replace the 
name $n$ of a student by a student id $i_n$, and then for each course
$c$ that is taken by $n$, one would like to include the tuple
$(i_n,c)$ in the table \enrollment. Unfortunately, as shown
in \cite{composing}, it is not possible to express this relationship
by using a set of st-tgds. In particular, a st-tgd of the form:
\begin{eqnarray}\label{eq-tgd}
\takes(n,c) & \rightarrow & \exists y \, \enrollment(y,c)
\end{eqnarray}
does not express the desired relationship, as it may associate a distinct
student id $y$ for each tuple $(n,c)$ in \takes\ and, thus, it may
create several identifiers for the same student name.  
\end{example} 


The previous example shows that in order to express the composition of mappings
specified by st-tgds, one has to use a language more expressive than
st-tgds. However, the example gives little information about what the
right language for composition is. In fact, the composition of
mappings $\M_{12}$ and $\M_{23}$ in this example can be defined in
first-order logic (FO):
\begin{eqnarray*}
\forall n \exists y \forall c \, (\takes(n,c) \to \enrollment(y,c)),
\end{eqnarray*}
which may lead to the conclusion that FO is a good alternative to
define the composition of mappings specified by st-tgds. However, a
complexity argument shows that this conclusion is wrong. More
specifically, given mappings $\M_{12} = (\R_1, \R_2, \Sigma_{12})$ and
$\M_{23} = (\R_2,
\R_3, \Sigma_{23})$, where $\Sigma_{12}$ and $\Sigma_{23}$ are sets of
st-tgds, define the {\em composition problem for $\M_{12}$ and
$\M_{23}$}, denoted by $\compp(\M_{12},\M_{23})$, as the problem of
verifying, given $I_1 \in \inst(\R_1)$ and $I_3 \in \inst(\R_3)$,
whether $(I_1,I_3) \in \M_{12} \circ \M_{23}$.  
If the composition of $\M_{12}$ with $\M_{23}$ is
defined by a set $\Sigma$ of formulas in some logic, then
$\compp(\M_{12},\M_{23})$ is reduced to the problem of verifying
whether a pair of instances $(I_1,I_3)$ satisfies $\Sigma$. In
particular, if $\Sigma$ is a set of FO formulas, then 
the complexity of $\compp(\M_{12},\M_{23})$ is in LOGSPACE, as the
complexity of verifying whether a fixed set of FO formulas is
satisfied by an instance is in LOGSPACE \cite{V82}. Thus, if for some mappings $\M_{12}$ and $\M_{23}$, the
complexity of the composition problem is higher than LOGSPACE, one can
conclude that FO is not capable of expressing the composition. In
fact, this higher complexity is proved in \cite{composing}.

\begin{theorem}[\cite{composing}]\label{theo-np-so-tgd}
For every pair of mappings $\M_{12}$, $\M_{23}$ specified by st-tgds,
$\compp(\M_{12},\M_{23})$ is in NP. Moreover, there exist mappings
$\M^\star_{12}$ and $\M^\star_{23}$ specified by st-tgds such 
that $\compp(\M^\star_{12},\M^\star_{23})$ is NP-complete.
\end{theorem}

Theorem \ref{theo-np-so-tgd} not only shows that FO is not the right
language to express the composition of mappings given by st-tgds, but
also gives a good insight on what needs to be added to st-tgds to
obtain a language closed under composition. Given that
$\compp(\M_{12},\M_{23})$ is in NP, we know by Fagin's Theorem that
the composition can be defined by an existential second-order logic
formula \cite{F74,L04}. In fact, Fagin et al. use this property in
\cite{composing} to obtain the right language for composition. More
specifically, Fagin et al. extend st-tgds with
existential second-order quantification, which gives rise to the class
of SO-tgds \cite{composing}. Formally, given schemas $\R_1$ and $\R_2$ with no relation
symbols in common, a {\em second-order tuple-generating dependency
from $\R_1$ to $\R_2$} (SO-tgd)
is a
formula of the form $\exists \bar f \, (\forall \bar x_1 (\varphi_1
\rightarrow \psi_1)
\wedge \cdots \wedge \forall \bar x_n (\varphi_n \rightarrow
\psi_n))$, 
where (1) each member of $\bar f$ is a function symbol, (2)
each formula $\varphi_i$ ($1 \leq i \leq n$) is a conjunction of
relational atoms of the form $S(y_1, \ldots, y_k)$ and equality atoms
of the form $t = t'$, where $S$ is a $k$-ary relation symbol of
$\R_1$ and $y_1$, $\ldots$, $y_k$ are (not necessarily distinct) variables in
$\bar x_i$, and $t$, $t'$ are terms built from $\bar x_i$ and $\bar
f$, (3) each formula $\psi_i$ ($1 \leq i \leq n$) is a conjunction of
relational atomic formulas over $\R_2$
mentioning terms built from $\bar x_i$ and $\bar f$, and (4) each
variable in $\bar x_i$ ($1 \leq i \leq n$) appears in some relational
atom of $\varphi_i$. 

In \cite{composing}, Fagin et al.~show that SO-tgds are the right
dependencies for expressing the composition of mappings given by
st-tgds. First, it is not difficult to see that every set of
st-tgds can be transformed into an SO-tgd. For example, set
$\Sigma_{12}$ from Example
\ref{exa-tgs-compo} is equivalent to the following SO-tgd: 
\begin{align*}
\exists f \bigg(&\forall n \forall c \, (\takes(n,c) \to \takes_1(n,c))
\ \wedge\\ 
&\forall n \forall c \, (\takes(n,c) \to \student(n,f(n,c)))\bigg).
\end{align*}
Second, Fagin et al.~show that SO-tgds are closed under composition.
\begin{theorem}[\cite{composing}]\label{comp-so-tgd}
Let $\M_{12}$ and $\M_{23}$ be mappings specified by SO-tgds. Then the
composition $\M_{12} \circ \M_{23}$ can also be specified by an
SO-tgd.
\end{theorem}
It should be noticed that the previous theorem can also be applied to
mappings that are specified by finite sets of SO-tgds, as these
dependencies are closed under conjunction. Moreover, it is important
to notice that Theorem \ref{comp-so-tgd} implies that the composition
of a finite number of mappings specified by st-tgds can be defined by
an SO-tgd, as every set of st-tgds can be expressed as an SO-tgd.

\begin{theorem}[\cite{composing}]\label{composing-theo}
The composition of a finite number of mappings, each defined by a
finite set of st-tgds, is defined by an SO-tgd. 
\end{theorem} 

\begin{example}\label{so-exa} 
Let $\M_{12}$ and $\M_{23}$ be the mappings defined in Example
\ref{exa-tgs-compo}. The following SO-tgd correctly specifies the
composition of these two mappings:
\begin{eqnarray*}
\exists g \bigg(\forall n \forall c \, (\takes(n,c) \to
\enrollment(g(n),c))\bigg). 
\end{eqnarray*}
\end{example}

\noindent
Third, Fagin et al.~prove in \cite{composing} that the converse of
Theorem \ref{composing-theo} also holds, thus showing that SO-tgds are exactly
the right language for representing the composition of mappings given
by st-tgds.
\begin{theorem}[\cite{composing}]
Every SO-tgd defines the composition of a finite number of mappings,
each defined by a finite set of st-tgds.   
\end{theorem}
Finally, Fagin et al.~in \cite{composing} also study the complexity of
composing schema mappings. More specifically, they provide an
exponential-time algorithm that given two mappings $\M_{12}$ and
$\M_{23}$, each specified by an SO-tgd, returns a mapping $\M_{13}$
specified by an SO-tgd and equivalent to the composition of $\M_{12}$
and $\M_{23}$. Furthermore, they show that exponentiality is unavoidable in
such an algorithm, as
there exist mappings $\M_{12}$ and $\M_{23}$, each specified by a
finite set of st-tgds, such that every SO-tgd that defines the
composition of $\M_{12}$ and $\M_{23}$ is of size exponential in the
size of $\M_{12}$ and $\M_{23}$.

In \cite{NBM07}, Nash et al.~also study the composition problem and
extend the results of \cite{composing}. In particular, they study the
composition of mappings given by dependencies that need not be
source-to-target, and for all the classes of mappings considered in
that paper, they provide an algorithm that attempts to compute the
composition and give sufficient conditions that guarantee that the
algorithm will succeed.

\subsection{Composition under closed world semantics}
\label{sec-comp-ext}
\noindent

In~\cite{L06}, Libkin proposes an alternative semantics for schema
mappings and, in particular, for data exchange.  Roughly speaking, the
main idea in~\cite{L06} is that when exchanging data with a set
$\Sigma$ of st-tgds and a source instance $I$, one generates a target
instance $J$ such that every tuple in $J$ is \emph{justified} by a
formula in $\Sigma$ and a set of tuples from $I$.  A target instance
$J$ that satisfies the above property is called a
\emph{closed-world solution} for $I$ under $\Sigma$~\cite{L06}.
In~\cite{LS08}, Libkin and Sirangelo propose the language of
$\sks$, that slightly extends the syntax of SO-tgds, and study the
composition problem under the closed-world semantics for mappings
given by sets of $\sks$.  Due to the lack of space, we do not give
here the formal definition of the closed-world semantics, but instead
we give an example that shows the intuition behind it 
(see~\cite{LS08} for a formal definition of the semantics
and of $\sks$).  

\begin{example}\label{cwa-exa}
Let $\sigma$ be the SO-tgd of Example~\ref{so-exa}.
Formula $\sigma$ is also a $\sk$~\cite{LS08}.
Consider now a source instance $I$ such that
$\takes^{I}=\{(\text{Chris},\text{logic})\}$,
and the instances $J_1$ and $J_2$ such that:
\begin{eqnarray*}
\enrollment^{J_1} & = & \{(075,\text{logic})\} \\
\enrollment^{J_2} & = & \{(075,\text{logic}),(084, \text{algebra})\}
\end{eqnarray*}
Notice that both $(I,J_1)$ and $(I,J_2)$ satisfy $\sigma$
(considering an interpretation for function $g$ such that
$g(\text{Chris}) = 075$). Thus, under the semantics based on logical
satisfaction~\cite{composing}, 
both $J_1$ and $J_2$ are solutions for $I$.
The crucial difference between $J_1$ and $J_2$ is that $J_2$ has an
\emph{unjustified} tuple~\cite{L06}; tuple $(075,\text{logic})$ is
\emph{justified} by tuple $(\text{Chris},\text{logic})$, while $(084,
\text{algebra})$ \emph{has no justification}. In fact, 
$J_1$ is a closed-world solution for $I$ under $\sigma$, but $J_2$ is
not~\cite{L06,LS08}.
\end{example}

Given a set $\Sigma$ of $\sks$ from $\R_1$ to $\R_2$, we
say that $\M$ is \emph{specified by $\Sigma$ under the closed-world semantics},
denoted by $\M=\cwa(\Sigma,\R_1,\R_2)$, if
$\M=\{(I,J)\mid I\in \inst(\R_1), J\in \inst(\R_2)$ and $J$ is a
closed-world solution for $I$ under $\Sigma\}$.
Notice that, as Example~\ref{cwa-exa} shows,
the mapping specified by a formula (or a set of formulas)
under the closed-world semantics
is different from the mapping specified by the same formula but under the semantics of~\cite{composing}.
Thus, it is not immediately clear whether a closure property
like the one in Theorem~\ref{comp-so-tgd} can be directly translated
to the closed-world semantics.
In this respect, Libkin and Sirangelo~\cite{LS08} show that the
language of $\sks$
is closed under composition.

\begin{theorem}[\cite{LS08}]
Let $\M_{12}=\cwa(\Sigma_{12},\R_1,\R_2)$ and
$\M_{23}$ $=$ $\cwa(\Sigma_{23},\R_2,\R_3)$,
where $\Sigma_{12}$ and $\Sigma_{23}$ are sets of $\sks$.
Then there exists a set $\Sigma_{13}$ of $\sks$ 
such that
$\M_{12}\comp \M_{23}=\cwa(\Sigma_{13},\R_1,\R_3)$.
\end{theorem}

\section{Inversion of Schema Mappings}
\label{sec-inv}
\noindent
In the recent years, the problem of inverting schema mappings has
received a lot of attention. In particular, the issue of providing a
{\em good} semantics for this operator turned out to be a difficult
problem.
Three main proposals for inverting mappings have been considered
so far in the literature:
\emph{Fagin-inverse}~\cite{inverse}, \emph{quasi-inverse}~\cite{quasi} and 
\emph{maximum recovery}~\cite{APR09}. 
In this section, we present and compare these 
approaches. 

Some of the notions mentioned above are only appropriate 
for certain classes of mappings.
In particular, the following two classes of mappings are
used in this section when defining and comparing 
inverses. A mapping
$\M$ from a schema $\R_1$ to a schema $\R_2$ is said to be {\em total}
if $\dom(\M) = \inst(\R_1)$, and is said to be {\em closed-down on the
left} if whenever $(I, J) \in \M$ and $I' \subseteq I$, it holds that
$(I', J) \in \M$.

Furthermore, whenever a mapping is specified by a set of formulas, we
consider source instances as just containing constants values, and
target instances as containing constants and null values.  This is a
natural assumption in a data exchange context, since target instances
generated as a result of exchanging data may be \emph{incomplete},
thus, null values are used as place-holders for unknown information.
In Section~\ref{sec-max-ext}, we consider inverses for alternative
semantics of mappings and, in particular, 
inverses for the \emph{extended semantics} that was proposed
in~\cite{FKPT09} to deal with incomplete information in source
instances. 

\subsection{Fagin-inverse and quasi-inverse}
\label{sec-fagin-quasi}
We start by considering the notion of inverse proposed by Fagin
in~\cite{inverse}, and that we call Fagin-inverse in this
paper%
\footnote{Fagin~\cite{inverse} named his notion just as \emph{inverse} of a schema mapping.
Since we are comparing different semantics for the \emph{inverse}
operator, we reserve the term \emph{inverse} to refer to this operator
in general, and use the name \emph{Fagin-inverse} for the notion
proposed in~\cite{inverse}.}.  
Roughly speaking, Fagin's definition is based on the idea that
a mapping composed with its inverse should be equal to the identity
schema mapping. Thus, given a schema $\R$, Fagin first defines an
\emph{identity mapping} $\IdF$ as $\{(I_1, I_2) \mid I_1, I_2$ are
instances of $\R$ and~$I_1 \subseteq I_2\}$.
Then a mapping $\M'$ is said to be a \emph{Fagin-inverse} of a mapping $\M$ if $\M
\circ \M' = \IdF$. 
Notice that $\IdF$ is not the usual identity relation over $\R$.
As explained in~\cite{inverse}, $\IdF$ is appropriate as 
an identity for mappings that are total and closed-down on the left
and, in particular,
for the class of mappings specified by st-tgds.

\begin{example}
Let $\M$ be a mapping specified by st-tgds $S(x) \to U(x)$ and $S(x)
\to V(x)$. 
Intuitively, $\M$ is Fagin-invertible since all the information in the
source relation $S$ is transferred to both relations $U$ and $V$ in the
target.  In fact, the mapping $\M'$ specified by ts-tgd $U(x) \to
S(x)$ is a Fagin-inverse of $\M$ since $\M \circ \M' =
\IdF$. Moreover,
the mapping $\M''$ specified by ts-tgd $V(x) \to S(x)$ is also
a Fagin-inverse of $\M$, which shows that there need not be a unique
Fagin-inverse.
\end{example}

A first fundamental question about any notion of inverse is for which
class of mappings is guaranteed to exist.
The following example from \cite{inverse} shows that Fagin-inverses
are not guaranteed to exist for mappings specified by st-tgds.

\begin{example}\label{exa-no-fag-inv}
Let $\M$ be a mapping specified by st-tgd $S(x,y) \to T(x)$. 
Intuitively, $\M$ has no Fagin-inverse since $\M$ only transfers the
information about the first component of $S$. 
In fact, it is formally proved in \cite{inverse} that this mapping is
not Fagin-invertible.
\end{example}


As pointed out in \cite{quasi}, the notion of Fagin-inverse is rather
restrictive as it is rare that a schema mapping possesses a
Fagin-inverse. Thus, there is a need for weaker
notions of inversion, which is the main motivation for
the introduction of the notion of quasi-inverse of a schema mapping in
\cite{quasi}.

The idea behind
quasi-inverses is to relax the notion of Fagin-inverse 
by
not differentiating between source instances that have the same space
of solutions. More precisely, let $\M$ be a mapping from a schema
$\R_1$ to a schema $\R_2$. Instances $I_{1}$ and $I_{2}$ of
$\R_1$ are \emph{data-exchange equivalent} w.r.t.~$\M$, denoted by
$I_{1} \sim_{\M}I_{2}$, if
$\Sol_{\M}(I_{1})=\Sol_{\M}(I_{2})$. 
For example, for the mapping $\M$ in Example~\ref{exa-no-fag-inv},
we have that $I_1\sim_\M I_2$, with $I_1 = \{S(1,2)\}$ and $I_2 = \{ S(1,3) \}$.
Then 
$\M'$ is said to be a quasi-inverse of $\M$ if the property 
$\M\comp \M' = \IdF$ holds \emph{modulo} the equivalence relation
$\sim_\M$. Formally, given a mapping
$\mathcal N$ from $\R$ to $\R$, mapping $\mathcal N[\sim_\M,\sim_\M]$ is
defined as 
\begin{multline*}
\{ (I_1,I_2) \in \inst(\R) \times \inst(\R) \mid \text{ exist }I_1',I_2'\text{ with} \\ 
I_1 \sim_\M I'_1,\ I_2 \sim_{\M} I'_2 \text{ and }
(I'_1,I'_2) \in \mathcal N\}
\end{multline*} 
Then a mapping $\M'$ is said to be a
\emph{quasi-inverse} of a mapping $\M$ if $(\M \circ
\M')[\sim_\M,\sim_\M] = \IdF[\sim_\M,\sim_\M]$.  

\begin{example}\label{exa-quasi}
Let $\M$ be a mapping specified by st-tgd $S(x,y) \to T(x)$. It was
shown in Example \ref{exa-no-fag-inv} that $\M$ does not have a
Fagin-inverse.
However, mapping $\M'$ specified by
ts-tgd $T(x) \to \exists y \, S(x,y)$
is a quasi-inverse of $\M$ \cite{quasi}.
Notice that for the source instance $I_1=\{S(1,2)\}$,
we have that $I_1$ and $I_2=\{S(1,3)\}$ are both solutions
for $I_1$ under the composition $\M\comp \M'$.
In fact, for every $I$ such that $I\sim_\M I_1$, we have that
$I$ is a solution for $I_1$ under $\M\comp \M'$.
\end{example}

In \cite{quasi}, the authors show that if a mapping $\M$
is Fagin-invertible, then a mapping $\M'$ is a Fagin-inverse of $\M$
if and only if $\M'$ is a quasi-inverse of $\M$.
Example \ref{exa-quasi} shows that the opposite direction does not hold. 
Thus, the notion of quasi-inverse is a strict generalization of the 
notion of Fagin-inverse.
Furthermore, the
author provides in \cite{quasi} a necessary and sufficient condition
for the existence of quasi-inverses for mappings specified by st-tgds,
and use this condition to show the following result:
\begin{proposition}[\cite{quasi}] There is a mapping $\M$ 
specified by a single st-tgd that has no quasi-inverse.
\end{proposition} 
Thus, although numerous non-Fagin-invertible schema mappings possess
natural and useful quasi-inverses \cite{quasi}, there are still simple
mappings specified by st-tgds that have no quasi-inverse. This leaves as
an open problem the issue of finding an appropriate notion of
inversion for st-tgds, and it is the main motivation for the introduction
of the notion of inversion discussed in the following section.

\subsection{Maximum recovery}
\label{sec-max}
We consider now the notion of maximum recovery introduced by Arenas et
al. in~\cite{APR08}. In that paper, the authors follow a different
approach to define a notion of inversion. In fact, the main goal of
\cite{APR08} is not to define a notion of inverse mapping, but instead
to give a formal definition for what it means for a mapping $\M'$ to
recover \emph{sound information} with respect to a mapping $\M$. Such a
mapping $\M'$ is called a recovery of $\M$ in \cite{APR08}. Given that,
in general, there may exist many possible recoveries for a given
mapping, Arenas et al.~introduce an order relation on recoveries
in \cite{APR08}, and show that this naturally gives rise to the notion
of maximum recovery, which is a mapping that brings back the maximum
amount of sound information.

Let $\M$ be a mapping from a schema $\R_1$ to a schema $\R_2$, and
$\IdM$ the identity schema mapping over $\R_1$, that is, $\IdM =
\{(I,I) \mid I \in \inst(\R_1)\}$. When trying to invert $\M$, the
ideal would be to find a mapping $\M'$ from $\R_2$ to $\R_1$ such that
$\M \circ \M' = \IdM.$ 
Unfortunately, in most cases this
ideal is impossible to reach (for example, for the case of mappings
specified by st-tgds \cite{inverse}). If for a mapping $\M$, there is
no mapping $\M_1$ such that $\M \circ \M_1 = \IdM$, at least one would
like to find a schema mapping $\M_2$ that does not forbid the
possibility of recovering the initial source data. This gives rise to
the notion of recovery proposed in \cite{APR08}. Formally, given
a mapping $\M$ from a schema $\R_1$ to a schema $\R_2$, a mapping
$\M'$ from $\R_2$ to $\R_1$ is a {\em recovery} of $\M$ if $(I,I)
\in \M \circ \M'$ for every instance $I \in \dom(\M)$~\cite{APR08}.

In
general, if $\M'$ is a recovery of $\M$, then the smaller the space of
solutions generated by $\M \circ \M'$, the more informative $\M'$ is
about the initial source instances. This naturally gives rise to the
notion of maximum recovery; given a mapping $\M$ and a recovery $\M'$
of it, $\M'$ is said to be a {\em maximum recovery} of $\M$ if for
every recovery $\M''$ of $\M$, it holds that $\M \circ \M' \subseteq
\M \circ \M''$ \cite{APR08}.

\begin{example}\label{exa-max}
In \cite{quasi}, it was shown that the schema mapping $\M$ specified
by st-tgd 
\[
E(x, z) \wedge E(z, y) \to F(x, y) \wedge M(z)
\] 
has neither a Fagin-inverse nor a quasi-inverse. 
However, it is possible to show
that the schema mapping $\M'$ specified by ts-tgds:
\begin{eqnarray*}
F(x, y) & \to & \exists u \, (E(x, u) \wedge E(u, y)),\\
M(z) & \to & \exists v \exists w \, (E(v, z) \wedge E(z,w)), 
\end{eqnarray*}
is a maximum recovery of $\M$.
Notice that, intuitively, the mapping $\M'$ is making the \emph{best effort}
to recover the initial data transferred by $\M$.
\end{example}

In \cite{APR08}, Arenas et al.~study the relationship between the
notions of Fagin-inverse, quasi-inverse and maximum recovery. It
should be noticed that the first two notions are only appropriate for
total and closed-down on the left mappings~\cite{inverse,APR08}.
Thus, the comparison in \cite{APR08} focus on these mappings. 
More precisely, it is
shown in \cite{APR08} that for every mapping $\M$ that is total and
closed-down on the left, if $\M$ is Fagin-invertible, then $\M'$ is a
Fagin-inverse of $\M$ if and only if $\M'$ is a maximum recovery of
$\M$. Thus, from Example \ref{exa-max}, one can conclude that the
notion of maximum recovery strictly generalizes the notion of
Fagin-inverse. 
The exact relationship between the
notions of quasi-inverse and maximum recovery is a bit more
involved. For every mapping $\M$ that is total and
closed-down on the left, it is shown in \cite{APR08} that if $\M$ is
quasi-invertible, then $\M$ has a maximum recovery and, furthermore,
every maximum recovery of $\M$ is also a quasi-inverse of $\M$.

In \cite{APR08}, the authors provide a 
necessary and sufficient
condition for the existence of a maximum recovery. It is important to
notice that this is general condition as it can be applied to any mapping, as long as it is defined as a set of pairs of instances.
This condition is used in \cite{APR08} to prove that every mapping
specified by a set of st-tgds has a maximum recovery.

\begin{theorem}[\cite{APR08}]\label{theo-max-exists}
Every mapping $\M$ specified by a finite set of st-tgds has a maximum
recovery.
\end{theorem}

\ignore{
Although one may be tempted to think that this is the end of the
story, there is an issue that was not considered by Arenas et al.~in
\cite{APR08}: the possible use of null values in source instances. It
is important to notice that the definition of maximum recovery also
applies to the case of mappings containing null values in source and
target instances, as this notion was defined for mappings viewed as
pair of instances without imposing any restrictions on the use of null
values
\cite{APR08}. However, as shown in the following section, the
application of this notion to the case of mappings specified by
st-tgds requires work if source instances are allowed to contain null
values. 
}

\subsection{Inverses for alternative semantics}
\label{sec-max-ext}

When mappings are specified by sets of logical formulas, we have considered
the usual semantics of mappings based on logical satisfaction.
However, some alternative semantics have been considered in the
literature, such as the \emph{closed world semantics}~\cite{L06},
the \emph{universal semantics}~\cite{de},
and the \emph{extended semantics}~\cite{FKPT09}. 
Although some of the notions of inverse discussed in the previous sections
can be directly applied to these alternative semantics, 
the positive and negative results on the existence of inverses 
need to be reconsidered in these particular cases.
In this section, we focus on this problem for the universal and extended
semantics of mappings.


\subsubsection{Universal solutions semantics}

Recall that a homomorphism from an instance $J_1$ to an instance $J_2$
is a function $h : \dom(J_1) \to \dom(J_2)$ such that (1) $h(c) = c$
for every constant $c \in \dom(J_1)$, and (2) for every fact $R(a_1, \ldots, a_k)$
in $J_1$, fact $R(h(a_1), \ldots, h(a_k))$ is in $J_2$.
Given a mapping $\M$ and a source instance $I$, a target instance
$J\in \Sol_\M(I)$ is a universal solution for $I$ under $\M$ if
for every $J'\in \Sol_\M(I)$, there exists a homomorphism from $J$ to $J'$.
It was shown in~\cite{de,core} that universal solutions have several desirable
properties for data exchange.
In view of this fact, an alternative semantics 
based on universal solutions was proposed in~\cite{core} 
for schema mappings.
Given a mapping $\M$, the mapping $u(\M)$ is defined
as the set of pairs 
\[
\{(I,J)\mid J\text{ is a universal solution for }I\text{ under }\M\}.
\]
Mapping $u(\M)$ was introduced in~\cite{core} in order to give a
clean semantics for answering target queries after exchanging data with mapping $\M$.
By combining the results on universal solutions for mappings given
by st-tgds in~\cite{de} and the results in~\cite{APR09} on the existence of maximum recoveries,
one can easily prove the following:

\begin{proposition}\label{univ-sol-semantic}
Let $\M$ be a mapping specified by a set of st-tgds.
Then $u(\M)$ has a maximum recovery. Moreover, the mapping
$(u(\M))^{-1} = \{ (J,I) \mid (I,J) \in u(\M)\}$ is a maximum recovery
of $u(\M)$. 
\end{proposition}

\subsubsection{Extended solutions semantics}

A more delicate issue regarding the semantics of mappings was
considered in~\cite{FKPT09}.  In this paper, Fagin et al.~made the
observation that almost all the literature about data exchange 
and, in particular, the literature about inverses of schema mappings,
assume that source instances do not have null values.  Since null
values in the source may naturally arise when using inverses of
mappings to exchange data, the authors relax the restriction on source
instances allowing them to contain values in $\C\cup \N$.  In fact,
the authors go a step further and propose new refined notions for
inverting mappings that consider nulls in the source.  In particular,
they propose the notions of \emph{extended inverse}, and of
\emph{extended recovery} and \emph{maximum extended recovery}.  In this section, we review the definitions of the latter two
notions and compare them with the previously proposed notions of
recovery and maximum recovery.

The first observation to make is that since null values are intended to
represent \emph{missing} or \emph{unknown} information, they
should not be treated naively as constants~\cite{IL84}.
In fact, as shown in~\cite{FKPT09}, if one treats nulls in that way,
the existence of a maximum recovery for mappings given by st-tgds is no longer guaranteed.

\begin{example}\label{no-max-rec}
Consider a source schema $\{S(\cdot,\cdot)\}$ 
where instances may contain null values,
and let $\M$ be a mapping specified by st-tgd
$S(x,y)  \rightarrow  \exists z \, (T(x,z) \wedge T(z,y))$.
Then $\M$ has no maximum recovery if one
considers a na\"{\i}ve semantics where null elements 
are used as constants in the source~\cite{FKPT09}.
\end{example}


Since nulls should not be treated naively when
exchanging data, in~\cite{FKPT09} the authors proposed a new
way to deal with null values.
Intuitively, the idea in~\cite{FKPT09} is to \emph{close} mappings
under homomorphisms. 
This idea is supported by the fact that nulls are intended
to represent unknown data, thus, it should be possible to replace them
by arbitrary values.
Formally, given a mapping $\M$, define $e(\M)$, the \emph{homomorphic extension}
of $\M$, as the mapping:
\begin{align*}
\{(I,J) \mid \ & \exists (I',J') \,:\, (I',J')\in \M \text{ and there exist}\\
&\text{homomorphisms from $I$ to $I'$ and from $J'$ to }J\}. 
\end{align*}
Thus, for a mapping $\M$ that has nulls in source and target instances,
one does not have to consider $\M$ but $e(\M)$ as the mapping
to deal with for exchanging data and computing mapping operators, since
$e(\M)$ treats nulls in a meaningful way~\cite{FKPT09}.
The following result shows that with this new semantics
one can avoid anomalies
as the one shown in Example~\ref{no-max-rec}.
\begin{theorem}[\cite{FKPT09b}]\label{ex-max-ex}
For every mapping $\M$ specified by a set of st-tgds and with nulls in
source and target instances, $e(\M)$ has a maximum recovery.
\end{theorem}
 
As mentioned above, Fagin et al.~go a step further in \cite{FKPT09} by
introducing new notions of inverse for mappings that consider nulls in
the source. More specifically, a mapping $\M'$ is said to be an
\emph{extended recovery} of $\M$  if 
$(I, I) \in e(\M) \circ e(\M')$, for every source instance~$I$.
Then given an extended recovery $\M'$ of $\M$, the mapping
$\M'$ is said to be a \emph{maximum extended recovery} of $\M$ 
if for every extended recovery $\M''$ of $\M$, it holds that
$e(\M) \circ e(\M') \subseteq e(\M) \circ e(\M'')$~\cite{FKPT09}.

At a first glance, one may think that the notions of maximum
recovery and maximum extended recovery are incomparable.
Nevertheless, the next result shows that there is a tight connection
between these two notions.
In particular, it shows that the notion proposed in~\cite{FKPT09}
can be defined in terms of the 
notion of maximum recovery.

\begin{theorem}\label{max-ext-rec-equiv}
A mapping $\M$ has a maximum extended recovery if and only if
$e(\M)$ has a maximum recovery.
Moreover, $\M'$ is a maximum extended recovery of $\M$ 
if and only if $e(\M')$ is a maximum recovery of $e(\M)$.
\end{theorem}
In \cite{FKPT09}, it is proved that every mapping specified by
a set of st-tgds and considering nulls in the source has a maximum
extended recovery. It should be noticed that this result is also
implied by Theorems \ref{ex-max-ex} and \ref{max-ext-rec-equiv}.

Finally, another conclusion that can be drawn from the above result is
that, all the machinery developed in~\cite{APR08,APR09} for the notion of
maximum recovery can be applied over maximum extended recoveries, and
the extended semantics for mappings,
thus giving a new insight about inverses of mappings with null values
in the source.

\subsection{Computing the inverse}
Up to this point, we have introduced and compared three notions of
inverse proposed in the literature, focusing mainly on the
fundamental problem of the existence of such inverses. In this
section, we study the problem of computing these inverses. More
specifically, we present some of the algorithms that have been
proposed in the literature for computing them, and we study the
languages used in these algorithms to express these inverses.

Arguably, the most important problem to solve in this area is the
problem of computing inverses of mappings specified by st-tgds. This
problem has been studied for the case of Fagin-inverse
\cite{quasi,FN08}, quasi-inverse \cite{quasi}, maximum recovery
\cite{APR08,APRR09,APR09} and maximum extended recovery
\cite{FKPT09,FKPT09b}. In this section, we start by presenting the
algorithm proposed in \cite{APR09} for computing maximum recoveries of
mappings specified by st-tgds, which by the results of Sections
\ref{sec-fagin-quasi} and \ref{sec-max} can also be used to compute
Fagin-inverses and quasi-inverses for this class of
mappings. Interestingly, this algorithm is based on {\em query
rewriting}, which greatly simplifies the process of computing such
inverses. 

Let $\M$ be a mapping from a schema $\R_1$ to a schema $\R_2$ and $Q$
a query over schema $\R_2$.  Then a query $Q'$ is said to be a
\emph{rewriting of $Q$ over the source} if $Q'$ is a query over $\R_1$
such that for every $I \in \inst(\R_1)$, it holds that $Q'(I) =
\certain_\M(Q,I)$.  That is, to obtain the set of certain answers of
$Q$ over $I$ under $\M$, one just has to evaluate its rewriting $Q'$
over instance $I$. 

The computation of a rewriting of a conjunctive query is a basic step
in the first algorithm presented in this section. This problem has
been extensively studied in the database area
\cite{LMSS95,LRO96,DG97,AD98,PH01} and, in particular, in the data
integration context \cite{Lev00,H01,integration}. The following
algorithm uses a query rewriting procedure {\sc QueryRewriting} to
compute a maximum recovery of a mapping $\M$ specified by a set
$\Sigma$ of st-tgds. In the algorithm, if $\x = (x_1, \ldots, x_k)$,
then $\pc(\x)$ is a shorthand for $\pc(x_1) \wedge \cdots \wedge
\pc(x_k)$.

\medskip\noindent
\underline{{\bf Algorithm} {\sc MaximumRecovery}($\M$)} \\
{\bf Input}: $\M =(\So,\Ta,\Sigma)$, where $\Sigma$ is a
set of st-tgds.\\  
{\bf Output}: $\M'=(\Ta,\So,\Sigma')$, where $\Sigma'$ is
a set of $\tu{\cqa{\pc}}{\ucqa{=}}$ ts-dependencies and $\M'$
is a maximum recovery of $\M$.

\begin{list}
{}
{
\setlength{\itemsep}{-2pt}
\setlength{\leftmargin}{0pt} 
\settowidth{\labelwidth}{1}
\setlength{\itemindent}{10pt}
\setlength{\topsep}{-5pt}
}
\item[\bf 1.] Start with $\Sigma'$ as the empty set.
\item[\bf 2.] For every dependency of the form
$\varphi(\x)\to \exists \y \, \psi(\x,\y)$ in $\Sigma$, do the following: 
  \begin{list}
  {}
  {
  \setlength{\itemsep}{-2pt}
  \setlength{\leftmargin}{10pt} 
  \settowidth{\labelwidth}{{1}}
  \setlength{\itemindent}{10pt}
  \setlength{\topsep}{-3pt}
  }  
  \item[\bf (a)] 
  Let $Q$ be the query defined by $\exists \y \, \psi(\x,\y)$.
  \item[\bf (b)] \label{step-rew}
  Use {\sc QueryRewriting}($\M,Q$) to compute a formula $\alpha(\x)$ in $\ucqa{=}$ 
  that is a rewriting of $\exists \y \, \psi(\x,\y)$ over the
source.    
  \item[\bf (c)] \label{step-add}
  Add dependency $\exists \y \, \psi(\x,\y) \wedge \C(\x) \to \alpha(\x)$ to $\Sigma'$. 
  \end{list}
\item[\bf 3.] Return $\M'=(\Ta,\So,\Sigma')$.
\hfill\mbox{}\qed
\end{list}
\medskip

\begin{theorem}[\cite{APR08,APR09}]
Let $\M=(\So,\Ta,\Sigma)$, where $\Sigma$ is a set of st-tgds.
Then {\sc MaximumRecovery}$(\M)$ computes a maximum recovery of $\M$
in exponential  time in the size of $\Sigma$, which is specified by a
set of $\tu{\cqa{\C}}{\ucqa{=}}$ dependencies. Moreover, if $\M$ is
Fagin-invertible (quasi-invertible), then the output of {\sc
MaximumRecovery}$(\M)$ is a Fagin-inverse (quasi-inverse) of $\M$.
\end{theorem}

It is important to notice that the algorithm {\sc MaximumRecovery}
returns a mapping that is a Fagin-inverse of an input mapping $\M$
whenever $\M$ is Fagin-invertible, but it does not check whether $\M$
indeed satisfies this condition (and likewise for the case of
quasi-inverse). 
In fact, it is not immediately clear whether the problem of checking
if a mapping given by a set of st-tgds has a Fagin-inverse is decidable.  
In~\cite{FN08}, the authors solve this problem showing the following:
\begin{theorem}[\cite{FN08}]
The problem of verifying whether a mapping specified by a set of
st-tgds is Fagin-invertible is coNP-complete.
\end{theorem}
Interestingly, it is not known whether the previous problem is
decidable for the case of the notion of quasi-inverse.

One of the interesting features of algorithm {\sc MaximumRecovery} is
the use of query rewriting, as it allows to reuse in the computation
of an inverse the large number of techniques developed to deal with
the problem of query rewriting. However, one can identify two
drawbacks in this procedure. First, algorithm {\sc
MaximumRecovery} returns a mapping that is specified by a set of
$\tu{\cqa{\C}}{\ucqa{=}}$ dependencies. Unfortunately, this type of
mappings are difficult to use in the data exchange context.  In
particular, it is not clear whether the standard chase procedure could
be used to produce a single canonical target database in this case,
thus making the process of exchanging data and answering queries much
more complicated. Second, the output mapping of {\sc
MaximumRecovery} can be of exponential size in the size of the input
mapping. Thus, a natural question at this point is whether simpler and
smaller inverse mappings can be computed. In the rest of this section,
we show some negative results in this respect, and also some efforts
to overcome these limitations by using more expressive mapping
languages.

The languages needed to express Fagin-inverses and quasi-inverses are
investigated in \cite{quasi,FN08}. In the respect, the first negative
result proved in
\cite{quasi} is that there exist quasi-invertible mappings specified
by st-tgds whose quasi-inverse cannot be specified by st-tgds. In
fact, it is proved in \cite{quasi} that the quasi-inverse of a mapping
given by st-tgds can be specified by using $\tu{\cqa{\neq,\pc}}{\ucq}$
dependencies, and that inequality, predicate $\pc(\cdot)$ and
disjunction are all unavoidable in this language in order to express
such quasi-inverse. For the case of Fagin-inverse, it is shown in
\cite{quasi} that disjunctions are not needed, that is, the
class of $\tu{\cqa{\neq,\pc}}{\cq}$ dependencies is expressive enough
to represent the Fagin-inverse of a Fagin-invertible mapping specified
by a set of st-tgds. In \cite{inverse,FN08}, it is proved a second negative
result about the languages needed to express Fagin-inverses, namely
that there is a family of Fagin-invertible mappings $\M$ specified by
st-tgds such that the size of every Fagin-inverse of $\M$ specified by
a set of $\tu{\cqa{\neq,\pc}}{\cq}$ dependencies is exponential in the
size of $\M$. Similar results are proved in \cite{APR08,APR09} for the
case of maximum recoveries of mappings specified by st-tgds. More specifically, it
is proved in \cite{APR08} that the maximum recovery of a mapping given
by st-tgds can be specified by using $\tu{\cqa{\pc}}{\ucqa{=}}$
dependencies, and that equality, predicate $\pc(\cdot)$ and
disjunction are all unavoidable in this language in order to express
such maximum recovery. Moreover, it is proved in \cite{APR09} that
there is a family of mappings $\M$ specified by st-tgds such that the
size of every maximum recovery of $\M$ specified by a set of
$\tu{\cqa{\pc}}{\ucqa{=}}$ dependencies is exponential in the size of
$\M$.

In view of the above negative results, Arenas et al.~explore in
\cite{APRR09} the possibility of using a more expressive language for
representing inverses. In particular, they explore the possibility of
using some extensions of the class of SO-tgds to express this operator.
In fact, Arenas et al.~provide in \cite{APRR09} a
polynomial-time algorithm that given a mapping $\M$ specified by a set
of st-tgds, returns a maximum recovery of $\M$, which is specified in
a language that extends SO-tgds (see \cite{APRR09} for a precise
definition of this language). 
It should be noticed that the algorithm presented in~\cite{APRR09}
was designed to compute maximum recoveries of mappings specified 
in languages beyond st-tgds, such as the language of \emph{nested mappings}~\cite{nested}
and plain SO-tgds (see Section \ref{sec-query} 
for a definition of the class of plain SO-tgds). 
Thus, the algorithm proposed in~\cite{APRR09}
can also be used to compute in polynomial time
Fagin-inverses (quasi-inverses) of Fagin-invertible (quasi-invertible)
mappings specified by st-tgds, nested mappings and plain SO-tgds.
Interestingly, a similar approach was
used in \cite{FKPT09b} to provide a polynomial-time algorithm for
computing the maximum extended recovery for the case
of mappings defined by st-tgds.

\section{Query-based notions of composition and inverse}
\label{sec-query}
As we have discussed in the previous sections, to express the
composition and the inverse of schema mappings given by st-tgds, one
usually needs mapping languages that are more expressive than
st-tgds, and that do not
have the same good properties for data exchange as st-tgds.

As a way to overcome this limitation, some weaker notions of
composition and inversion have been proposed in the recent years,
which are based on the idea that in practice one may be interested in
querying exchanged data by using only a particular class of
queries. In this section, we review these notions.


\subsection{A query-based notion of composition}

In this section, we study the notion of \emph{composition
w.r.t.~conjunctive queries} ($\cq$-composition for short) introduced
by Madhavan and Halevy~\cite{MH03}.
This semantics for composition can be defined in terms of the notion of
\emph{conjunctive-query equivalence} of mappings that was introduced in~\cite{MH03} 
for studying $\cq$-composition and generalized in~\cite{FKNP08} when studying 
optimization of schema mappings.
Two mappings $\M$ and $\M'$ from $\So$ to $\Ta$ are said to be
\emph{equivalent w.r.t.~conjunctive queries}, 
denoted by $\M\equiv_\cq\M'$, 
if for every conjunctive query $Q$,
the set of certain answers of $Q$ under $\M$ 
coincides with the set of certain answers of $Q$ under $\M'$.
Formally, $\M\equiv_{\cq} \M'$ if for every 
conjunctive query $Q$ over $\Ta$ and every instance $I$ of $\So$, 
it holds that $\certain_{\M}(Q, I) = \certain_{\M'}(Q, I)$.
Then $\cq$-composition can be defined as follows:
$\M_3$ is a $\cq$-composition of $\M_1$ and $\M_2$
if $\M_3\equiv_\cq \M_1\comp \M_2$.


A fundamental question about the notion of $\cq$-composition is
whether the class of st-tgds is closed under this notion.
This problem was implicitly studied by Fagin et al.~\cite{FKNP08}
in the context of schema mapping optimization.
In~\cite{FKNP08}, 
the authors consider the problem of whether a mapping
specified by an SO-tgd is $\cq$-equivalent to a mapping specified by st-tgds.
Thus, given that the composition of a finite number of mappings given
by st-tgds can be defined by an SO-tgd \cite{composing}, the latter
problem is a reformulation of the problem of testing whether st-tgds
are closed under $\cq$-composition. In fact, by using the results and
the examples in~\cite{FKNP08}, one can easily construct mappings
$\M_1$ and $\M_2$ given by st-tgds such that the $\cq$-composition of
$\M_1$ and $\M_2$ is not definable by a finite set of st-tgds. 

A second fundamental question about the notion of $\cq$-composition is
what is the right language to express it. Although this problem is
still open, in the rest of this section we shed light on this issue.
By the results in~\cite{composing}, we know that the language of SO-tgds
is enough to represent the $\cq$-composition of st-tgds. However, as
motivated by the following example, some features of SO-tgds are not
needed to express the $\cq$-composition of mappings given by st-tgds.

\newcommand{\emp}{\text{\tt Emp}}
\newcommand{\mgr}{\text{\tt Mgr}}
\newcommand{\selfmgr}{\text{\tt SelfMgr}}

\begin{example}{\bf \ (from \cite{composing})}
Consider a schema $\R_1$ consisting of one unary relation \emp\ 
that stores employee names, a schema $\R_2$ consisting of a binary 
relation \mgr$_{\text{\tt 1}}$ that assigns a manager to each employee,
and a schema $\R_3$ consisting of a binary relation \mgr\ 
intended to be a copy of {\tt Mgr}$_1$ and of a unary relation \selfmgr, 
that stores employees that are manager of themselves.
Consider now mappings $\M_{12}$ and $\M_{23}$ specified by the following sets of st-tgds:
\begin{eqnarray*}
\Sigma_{12} & = & \{\ \emp(e) \ \to \ \exists m\ \mgr_{\text{\tt 1}}(e,m) \ \}, \\
\Sigma_{23} & = & \{\ \mgr_{\text{\tt 1}}(e,m) \ \to \ \mgr(e,m), \\ 
& & \phantom{\{\ }\mgr_{\text{\tt 1}}(e,e) \ \to \ \selfmgr(e)\ \}. 
\end{eqnarray*}
Mapping $\M_{12}$ intuitively states that every employee must be associated with a manager.
Mapping $\M_{23}$ requires that a copy of every tuple in $\mgr_\text{\tt 1}$ must exists
in \mgr, and creates a tuple in \selfmgr\ whenever an employee is the manager of her/himself. 
It was shown in~\cite{composing} that the mapping $\M_{13}$ given by 
the following SO-tgd:
\begin{multline} 
\label{so-eq-map}
\exists f \big( \forall e (\emp(e)\ \to\  \mgr(e,f(e))) \wedge\\
\forall e(\emp(e) \wedge e = f(e) \rightarrow \selfmgr(e))\big) 
\end{multline}
represents the composition $\M_{12} \circ \M_{23}$.
Moreover, the authors prove in~\cite{composing} that the equality in the above
formula is strictly necessary to represent that composition. 
However, it is not difficult to prove that 
the mapping $\M_{13}'$ given by the following formula:
\begin{equation}\label{plain}
\exists f \big( \forall e (\emp(e) \ \to \ \mgr(e,f(e))) \big)
\end{equation}
is $\cq$-equivalent to $\M_{13}$, and thus, $\M_{13}'$ is a $\cq$-composition
of $\M_{12}$ and $\M_{23}$.
\end{example}
We say that formula~\eqref{plain} is a {\em plain SO-tgd}.
Formally, a plain SO-tgd from $\R_1$ to $\R_2$ is an SO-tgd satisfying the
following restrictions: (1) equality atoms are not allowed,
and (2) nesting of functions is not allowed. 
Notice that, just as SO-tgds, this language
is closed under conjunction and, thus, we talk about a mapping
specified by a plain SO-tgd (instead of a set of plain SO-tgds). 
The following result shows that even though the language of plain SO-tgds
is less expressive than the language of SO-tgds, they are equally expressive
in terms of $\cq$-equivalence.


\begin{lemma}\label{lemma-equiv-plain}
For every SO-tgd $\sigma$, there exists a plain SO-tgd $\sigma'$ such
that $\sigma \equiv_{\cq} \sigma'$.
\end{lemma}

It is easy to see that every
mapping specified by a set of st-tgds can be specified with a plain
SO-tgd. Moreover, the following theorem
shows that this language is closed under $\cq$-composition, thus
showing that this class 
of dependencies has good properties 
within the framework of $\cq$-equivalence. 
\begin{theorem} \label{theo-closed-compo-sod}
Let $\M_{12}$ and $\M_{23}$ be mappings specified by plain SO-tgds.
Then the $\cq$-composition of $\M_{12}$ and $\M_{23}$ can be specified
with a plain SO-tgd.
\end{theorem}
Thus, the $\cq$-composition of a finite number of mappings, each
specified by a set of st-tgds, is definable by a plain SO-tgd. It
should be noticed that Theorem \ref{theo-closed-compo-sod} is a
consequence of Lemma \ref{lemma-equiv-plain} and the fact that the
class of SO-tgds is closed under composition \cite{composing}.

Besides the above mentioned results, the language of plain SO-tgds
also has good properties regarding inversion. In particular, it is
proved in \cite{APRR09} that every plain SO-tgd has a maximum recovery, and,
moreover, it is given in that paper a polynomial-time algorithm to
compute it.  Thus, it can be argued that this
class of dependencies is more suitable for inversion than SO-tgds, as
there exist SO-tgds that do not admit maximum recoveries.



\subsection{A query-based notion of inverse}


In~\cite{APRR09}, the authors propose an alternative notion of inverse
by focusing on conjunctive queries. 
In particular, the authors first define the notion of \emph{$\cq$-recovery} as follows. A mapping $\M'$ is a $\cq$-recovery of $\M$ if for every
instance $I$ and conjunctive query $Q$, it holds that 
\begin{eqnarray*}
\certain_{\M\comp \M'}(Q,I) & \subseteq & Q(I). 
\end{eqnarray*}
Intuitively, this equation states that 
$\M'$ \emph{recovers sound information} for $\M$
w.r.t.~conjunctive queries since for every instance $I$, by posing a
conjunctive query $Q$
against the space of solutions for $I$ under $\M \circ \M'$, one can
only recover data that is already in the evaluation of $Q$ over $I$.
A \emph{$\cq$-maximum recovery} is then defined as a mapping that
recovers the maximum amount of sound information w.r.t.~conjunctive
queries.
Formally, a $\cq$-recovery $\M'$ of $\M$ is a $\cq$-maximum recovery
of $\M$ if for every other $\cq$-recovery $\M''$ of $\M$, it holds that
\begin{eqnarray*}
\certain_{\M\comp \M''}(Q,I) & \subseteq & \certain_{\M\comp \M'}(Q,I),
\end{eqnarray*} 
for every instance $I$ and conjunctive query $Q$.

In~\cite{APRR09}, the authors study several properties
about $\cq$-maximum recoveries.
In particular, they provide an algorithm to compute 
$\cq$-maximum recoveries for st-tgds showing the following:
\begin{theorem}[\cite{APRR09}]
Every mapping specified by a set of st-tgds has a $\cq$-maximum
recovery, which is
specified by a set of $\tu{\cq^{\pc,\neq}}{\cq}$ dependencies.
\end{theorem}
Notice that the language needed to express $\cq$-maximum recoveries of st-tgds
has the same good properties as st-tgds for data exchange.
In particular, the language is \emph{chaseable} in the sense that the 
standard chase procedure can be used to obtain a canonical solution.
Thus, compared to the notions of Fagin-inverse, quasi-inverse, and maximum recovery,
the notion of $\cq$-maximum recovery has two advantages:
(1) every mapping specified by st-tgds has a $\cq$-maximum recovery (which is not
the case for Fagin-inverses and quasi-inverses), and
(2) such recovery can be specified in a mapping language with
good properties for data exchange (which is not the case for quasi-inverses and
maximum recovery).

In~\cite{APRR09}, the authors also study the minimality of the language
used to express $\cq$-maximum recoveries, showing that inequalities and 
predicate $\pc(\cdot)$ are both needed to express the $\cq$-maximum recoveries
of mappings specified by st-tgds.

\section{Future Work}
\label{sec-con}
\ignore{
The importance of schema mappings has been increasing systematically
in the past years, since it is tied to the never ending proliferation
of systems that perform data-interoperability tasks.
Nowadays, the community recognizes
the need to develop techniques to manipulate these mappings specifications.
Of all the possible alternatives, the composition and inverse
operators have been
pointed out as a good place to start, for they
are the building blocks required to perform more complicated operations.

Nevertheless,
the definition of the appropriate semantics for the inverse and
composition operators has proven to be a non-trivial task, in which many
difficult decisions had to be taken.
In fact, the answer to each of these decisions has given rise to
different semantics
for the composition and inverse operator. Do we want operators that
are guaranteed to be consistent in a general scenario, or do we settle
for notions that allow only conjunctive query answering?
Certainly, the latter question involves a tradeoff, since the more
general operators
usually require more expressive languages, and their computation is
more complex.
On the other hand, a parallel question originates with the spread of
new semantics for
the schema mappings themselves. Under these new semantics, the
previously defined
operators have to be re-studied. This raises the importance of general
operators that are not tied to a particular schema mapping semantic,
such as the maximum recovery.
}

As many information-system problems involve not only the design and
integration of complex application artifacts, but also their
subsequent manipulation, the definition and implementation of some
operators for metadata management has been identified as a
fundamental issue to be solved \cite{B03b}. In particular, composition
and inverse have been identified as two of the fundamental operators
to be studied in this area, as they can serve as building blocks of
many other operators \cite{Melnik04,MBHR05}. In this paper, we have
presented some of the results that have been obtained in the recent
years about the composition and inversion of schema mappings.

Many problems remain open in this area. 
Up to now, XML schema mapping
languages have been proposed and studied~\cite{AL08, ALM09, TBM09},
but little attention has been paid to the formal study of XML schema
mapping operators.  For the case of composition, a first insight
has been given in~\cite{ALM09}, showing that the previous results
for the relational model are not directly applicable over
XML.  Inversion of XML schema mappings remains an unexplored field.

Regarding the relational model,
we believe that the future effort has to be focused in providing a
unifying framework for these operators, one that permits the
successful application of them.  A natural question,
for instance, is whether there exists a schema mapping language that
is closed under both composition and inverse.  Needless to say, this
unified framework will permit the modeling of more complex algebraic
operators for schema mappings.

\ignore{
One still open problem is to operate mappings between
XML schemas. 
Although XML-schema mapping languages have been proposed and
studied~\cite{AL08, ALM09, TBM09} little attention has been
paid to the formal study of XML-schema mapping operators.
in~\cite{AL08, ALM09},
For the case of the composition, a first insight has been given in~\cite{ALM09},
showing that the previous results obtained for the relational model
are not directly applicable over XML.
Invertion of XML-schema mappings remains an unexplored field.

Beside studying invertion and composition for database models
beyond the relational model,
we believe that the future effort has to be focused in providing a
unifying framework for these operators, one that permits the
successful application of them.  A natural question,
for instance, is whether there exists a schema mapping language that
is closed under both composition and inverse.  Needless to say, this
unified framework will permit the modeling of more complex algebraic
operations for schema mappings.}

\section*{Acknowledgments}
We would like to thank L. Libkin and R. Fagin for many useful
comments.  The authors were supported by: Arenas - Fondecyt grant
1090565; P\'erez - Conicyt Ph.D. Scholarship.

\newpage

\onecolumn 

\appendix

\section{Proofs and Intermediate Results}
\label{sec-app}

In this section, we provide proofs for the new results reported in
this survey.  Some of these proofs are related with the notion of
maximum recovery proposed in~\cite{APR08}, and its relationship with
some other notions of inverse.  The main tool used in this section
regarding maximum recoveries is described in the following
proposition.

\begin{proposition}[\cite{APR08}]\label{maxrec-carac}
$\M'$ is a maximum recovery of $\M$ if and only if $\M'$ is a recovery
of $\M$ and for every $(I_1,I_2)\in \M \comp \M'$, it holds that
$\Sol_{\M}(I_{2})\subseteq \Sol_{\M}(I_{1})$.  
\end{proposition}
It should be noticed that the above is a characterization of the
notion of maximum recovery for a mapping $\M$ that is total, that is,
if $\M$ is a mapping from $\R_1$ to $\R_2$, then
$\dom(\M)=\inst(\R_1)$.\footnote{In~\cite{APR08}, the authors provide
more general characterizations for mappings that are not necessarily
total by considering the notion of reduced recovery.}

\subsection{Proof of Proposition~\ref{univ-sol-semantic}}
\medskip

\noindent
Let $\M$ be a mapping specified by a set of st-tgds. We know by
\cite{de} that every source instance has a universal solution under
$\M$, and, thus, $u(\M)$ is a total mapping.  Next we show that $u(\M)$
and $u(\M)^{-1}$ satisfy the condition of
Proposition~\ref{maxrec-carac}, which implies that $u(\M)^{-1}$ is a
maximum recovery of $u(\M)$. 

It is straightforward to show that $u(\M)^{-1}$ is a recovery of
$u(\M)$ and, hence, it only remains to prove that for every tuple
$(I_1, I_2) \in u(\M) \circ u(\M)^{-1}$, it holds that: 
\begin{eqnarray*}
\Sol_{u(\M)}(I_2) & \subseteq & \Sol_{u(\M)}(I_1). 	
\end{eqnarray*}
Assume that $(I_1, I_2) \in u(\M) \circ u(\M)^{-1}$.  Then there
exists a target instance $J$ such that $(I_1,J) \in u(\M)$ and
$(J,I_2) \in u(\M)^{-1}$.  Thus, given that every solution in $u(\M)$
is a universal solution in $\M$, we have that $J$ is a universal
solution for both instances $I_1$ and $I_2$ in $\M$. Hence, we have by
Proposition 2.6 in \cite{de} that $\Sol_{\M}(I_1) = \Sol_{\M}(I_2)$
and, thus, $\Sol_{u(\M)}(I_2) \subseteq \Sol_{u(\M)}(I_1)$, which was
to be shown.

\subsection*{Proof of Proposition~\ref{max-ext-rec-equiv}}

We first introduce some notation to simplify the exposition.
Let $I_1$ and $I_2$ be instances of the same schema $\R$ with values in $\C \cup \N$.  
Recall that a \emph{homomorphism} from $I_1$ to $I_2$ is a function 
$h : \dom(I_1) \rightarrow \dom(I_2)$ such that, for every constant
value $a \in \C$, it holds that $h(a)=a$, and for every $R \in \R$ and
every tuple $(a_1, \ldots, a_k) \in R^{I_1}$, it holds 
$(h(a_1), \ldots, h(a_k)) \in R^{I_2}$. 
Consider a binary relation $\to$ defined as follows:
\begin{eqnarray*}
\to &  = & \{(I_1, I_2) \mid  \text{there exists a homomorphism from }
I_1\text{ to }I_2\}. 
\end{eqnarray*}
In~\cite{FKPT09}, relation $\to$ was introduced to simplify the definition
of the extended semantics of a mapping. In fact, given a mapping $\M$,
we have that 
\begin{eqnarray*}
e(\M) & = & \to\comp\ \M\ \comp \to.
\end{eqnarray*}
Notice that the relation $\to$ is \emph{idempotent}, that is, it holds
that $(\to\comp \to) \ = \ \to$.  In particular, we have that 
\begin{eqnarray}
\to \comp\ e(\M) & = & e(\M), \label{lidemp} \\
e(\M)\ \comp\to\ & = & e(\M). \label{ridemp}
\end{eqnarray}
Thus, if $I_1$, $I_2$, $J$ are instances such that $(I_1,I_2) \in \ \to$
and $(I_2,J)\in e(\M)$, then $(I_1,J)\in e(\M)$.  Hence, if $(I_1,I_2)
\in \ \to$, then it holds that $\Sol_{e(\M)}(I_2)\subseteq
\Sol_{e(\M)}(I_1)$.  We use this property in this proof.

Before proving the proposition, we make an additional observation. The
extended recovery of a mapping $\M$ is defined in~\cite{FKPT09} only
for the case when the domain of $e(\M)$ is the set of all source
instances.  More precisely, a mapping $\M'$ is said to be an extended
recovery of $\M$ in \cite{FKPT09} if \emph{for every source instance
$I$}, it holds that $(I,I)\in e(\M)\comp e(\M')$. Thus, it is only
meaningful to compare the notions of (maximum) extended recovery and
(maximum) recovery for the class of mappings $\M$ such that $e(\M)$ is
the set of all source instances. For this reason, if $\M$ is a mapping from
a schema $\R_1$ to a schema $\R_2$, then we assume in this proof that
$\dom(e(\M)) = \inst(\R_1)$. It should be noticed that this implies by
Proposition~\ref{maxrec-carac} that:
\begin{center}
$\M'$ is a maximum recovery of $e(\M)$\\
if and only if\\ 
$\M'$ is a recovery of $e(\M)$ and for every $(I_1,I_2)\in e(\M)\comp
\M'$, it holds that $\Sol_{e(\M)}(I_2)\subseteq \Sol_{e(\M)}(I_1)$. 
\end{center}  
We extensively use this property in this proof.

\medskip


Now we are ready to prove Proposition \ref{max-ext-rec-equiv}.  Let
$\M$ be a mapping from a schema $\So$ to a schema $\Ta$, and assume
that source instances are composed by null and constant values.  We
first show that $e(\M)$ has a maximum recovery if and only if $\M$ has
a maximum extended recovery.

\medskip

($\Rightarrow$) Assume that $e(\M)$ has a maximum recovery, and let
$\M'$ be a maximum recovery of $e(\M)$.  We show next that $\M'$ is
also a maximum extended recovery of $\M$.  Since $\M'$ is a recovery
of $e(\M)$, we have that $(I, I) \in e(\M) \circ \M'$ for every
instance $I$ of $\So$.  Moreover, from~\eqref{ridemp} we have that
$e(\M) \circ \M' = e(\M)\ \comp\to \comp\ \M'$ and, thus, $(I, I) \in
e(\M)\ \comp\to \comp\ \M'$ for every instance $I$ of $\So$.  Thus,
given that $(I, I) \in\ \to\ $ for every instance $I$ of $\So$, we
obtain that $(I, I) \in e(\M)\ 
\comp\to \comp\ \M'\ \comp\to\ = e(\M)\comp e(\M')$ for every instance
$I$ of $\So$, which implies that $\M'$ is an extended recovery of
$\M$. 

Now, let $\M''$ be an extended recovery of $\M$.  Then, as above, we
obtain that $(I, I) \in e(\M) \comp e(\M'')$ for every instance $I$
of $\So$. Thus, we have that $e(\M'')$ is a recovery of $e(\M)$.
Recall that $\M'$ is a maximum recovery of $e(\M)$ and, hence, we have
that $e(\M)\comp \M'\subseteq e(\M)\comp e(\M'')$, which implies that
$e(\M)\comp \M'\comp \to\ \subseteq e(\M)\comp e(\M'')\ \comp \to$.
Therefore, given that $e(\M)=e(\M)\ \comp \to$ and
$e(\M'')\ \comp \to\ =e(\M'')$ by \eqref{ridemp}, we have that $e(\M)\
\comp \to \comp\ \M'\ \comp \to\ \subseteq e(\M)\comp e(\M'')$, which
implies that $e(\M) \comp e(\M') \subseteq e(\M)\comp e(\M'')$.  Thus,
we have shown that $\M'$ is an extended recovery of $\M$, and that for
every other extended recovery $\M''$ of $\M$, it holds that $e(\M)\comp
e(\M')\subseteq e(\M)\comp e(\M'')$, which implies that $\M'$ is a
maximum extended recovery of $\M$.

\medskip

($\Leftarrow$) Now assume that $\M$ has a maximum extended recovery,
and let $\M'$ be a maximum extended recovery of $\M$.  Next we show
that $e(\M')$ is a maximum recovery of $e(\M)$.  

Given that $\M'$ is an extended recovery of $\M$, we have that
$(I,I)\in e(\M)\comp e(\M')$ for every instance $I$ of $\So$, which
implies that $e(\M')$ is a recovery of $e(\M)$. Thus, by
Proposition~\ref{maxrec-carac}, to prove that $e(\M')$ is a maximum
recovery of $e(\M)$, it is enough to show that
$\Sol_{e(\M)}(I_{2})\subseteq \Sol_{e(\M)}(I_{1})$ for every
$(I_1,I_2) \in e(\M) \circ e(\M')$.  Let $(I_1,I_2) \in e(\M) \circ
e(\M')$.  To prove that $\Sol_{e(\M)}(I_{2})\subseteq
\Sol_{e(\M)}(I_{1})$, we make use of the following mapping $\M^{\star}$
from $\Ta$ to $\So$:
\begin{eqnarray*}
\M^{\star} & = & \{(J,I) \mid I \text{ is an instance of } \So \text{
and } (I_1,J)\notin e(\M) \} \ \cup \\
&   & \{(J,I) \mid 
(I_1,J) \in e(\M) \text{ and } \Sol_{e(\M)}(I) \subseteq \Sol_{e(\M)}(I_{1})\}.
\end{eqnarray*}
We show first that $\M^\star$ is an extended recovery of $\M$, that
is, we show that for every instance $I$ of $\So$, it holds that $(I,I)\in
e(\M)\comp e(\M^\star)$.  First, assume that $\Sol_{e(\M)}(I)\subseteq
\Sol_{e(\M)}(I_1)$, and consider an arbitrary instance $J^\star$ such
that $(I,J^\star)\in e(\M)$.  Notice that $(I_1,J^\star)\in e(\M)$
since $\Sol_{e(\M)}(I)\subseteq \Sol_{e(\M)}(I_1)$.  Thus, we have
that $(J^\star,I)\in \M^\star$ and, hence, $(J^\star,I)\in
e(\M^\star)$. Therefore, given that $(I,J^\star)\in e(\M)$ and
$(J^\star,I)\in e(\M^\star)$, we conclude that $(I,I)\in 
e(\M)\comp e(\M^\star)$. Second, assume that $\Sol_{e(\M)}(I) \not\subseteq
\Sol_{e(\M)}(I_1)$.  Then there exists an instance $J^\star$ such
that $(I,J^\star)\in e(\M)$ and $(I_1,J^\star)\notin e(\M)$.  By
definition of $\M^\star$, we have that $(J^\star,I)\in \M^\star$
and, thus, $(J^\star,I)\in e(\M^\star)$. Thus, we also conclude that
$(I,I)\in e(\M)\comp e(\M^\star)$ in this case.

We are now ready to prove that for every $(I_1,I_2) \in e(\M) \circ
e(\M')$, it holds that $\Sol_{e(\M)}(I_{2})\subseteq
\Sol_{e(\M)}(I_{1})$.  Let $(I_1,I_2) \in e(\M) \circ e(\M')$.  Given that
$\M'$ is a maximum extended recovery of $\M$ and $\M^\star$ is an
extended recovery of $\M$, we have that $e(\M) \circ e(\M') \subseteq
e(\M) \circ e(\M^{\star})$ and, therefore, $(I_1, I_2) \in e(\M) \circ
e(\M^{\star})$. Thus, given that $e(\M) \circ e(\M^\star)=e(\M)\comp
\M^\star \circ \to$ by \eqref{ridemp}, we
conclude that there exist instances $J$ of $\Ta$ and $I_{2}'$ of $\So$
such that $(I_1,J)\in e(\M)$, $(J,I_2')\in \M^\star$ and $(I_{2}',
I_2) \in\ \rightarrow$. Hence, by definition of $\M^{\star}$, we have
that $\Sol_{e(\M)}(I_{2}') \subseteq \Sol_{e(\M)}(I_{1})$ (since
$(I_1,J) \in e(\M)$). But we also have that
$\Sol_{e(\M)}(I_{2})\subseteq \Sol_{e(\M)}(I_{2}')$ since $(I_{2}',
I_2) \in\ \rightarrow$, and, therefore, we conclude that
$\Sol_{e(\M)}(I_{2}) \subseteq \Sol_{e(\M)}(I_{1})$, which was to be
shown.

\medskip

Up to this point, we have shown that $e(\M)$ has a maximum recovery if
and only if $\M$ has a maximum extended recovery. In fact, from the
preceding proof, we conclude that:
\begin{enumerate}
\item[(a)] if $e(\M)$ has a maximum recovery $\M'$, then $\M'$ is a
maximum extended recovery of $\M$, and 

\item[(b)] if $\M$ has a maximum extended recovery $\M'$, then
$e(\M')$ is a maximum recovery of $e(\M)$. 
\end{enumerate} 
Next we prove the second part of Proposition~\ref{max-ext-rec-equiv},
that is, we prove that a mapping $\M'$ is a maximum extended recovery
of $\M$ if and only if $e(\M')$ is a maximum recovery of $e(\M)$. It
should be noticed that the ``only if'' direction corresponds to
property (b) above and, thus, we only need to show that if $e(\M')$ is
a maximum recovery of $e(\M)$, then $\M'$ is a maximum extended
recovery of $\M$.  

Assume that $e(\M')$ is a maximum recovery of $e(\M)$.  Then we have
that $e(\M')$ is a recovery of $e(\M)$ and, thus, $\M'$ is an extended
recovery of $\M$.  Now let $\M''$ be an extended recovery of $\M$.
Then we have that $e(\M'')$ is a recovery of $e(\M)$ and, hence,
$e(\M) \comp e(\M') \subseteq e(\M) \comp e(\M'')$ since $e(\M')$ is a
maximum recovery of $e(\M)$. Therefore, we conclude that $\M'$ is an
extended recovery of $\M$, and for every extended recovery $\M''$ of
$\M$, it holds that $e(\M)\comp e(\M')\subseteq e(\M)\comp e(\M'')$,
which means that $\M'$ is a maximum extended recovery of $\M$.  This
completes the proof of the proposition.

\end{document}